\documentclass[twocolumn,prb]{revtex4}

\usepackage{graphicx}

\newcommand{\BEQ}{\begin{equation}}
\newcommand{\EEQ}{\end{equation}}
\newcommand{\BEA}{\begin{eqnarray}}
\newcommand{\EEA}{\end{eqnarray}}

\begin{document}
\title{Stable Solution of the Simplest Spin Model for Inverse Freezing}

\date{\today}

\author{Andrea Crisanti$^\star$ and Luca Leuzzi$^\star$$^\diamond$}

\affiliation{ $\star$ Department of Physics, University of Rome, ``La
Sapienza'', Piazzale A. Moro 2, 00185, Rome, Italy 
\\ 
$\diamond$ SMC-INFM and Institute of Complex Systems (ISC) CNR, Via dei
Taurini 19, Rome, Italy}

\begin{abstract}
  We analyze the Blume-Emery-Griffiths model with disordered magnetic
  interaction that displays the inverse freezing
  phenomenon. The behavior of this spin-$1$ model in crystal field is
  studied throughout the phase diagram and the transition and spinodal
  lines for the model are computed using the Full Replica Symmetry
  Breaking Ansatz that always yields a thermodynamically stable phase.
  We compare the results both with the formulation of the same model
  in terms of Ising spins on lattice gas, where no reentrance takes
  place, and with the model with generalized spin variables recently
  introduced by Schupper and Shnerb [Phys.  Rev. Lett. {\bf 93} 037202
  (2004)], for which the reentrance is enhanced as the ratio between
  the degeneracy of full to empty sites increases. The simplest
  version of all these models, known as the Ghatak-Sherrington model,
  turns out to hold all the general features characterizing an inverse
  transition to an amorphous phase, including the right thermodynamic behavior.
\end{abstract}

\maketitle

%%%%%%%%%%%%%%%%%%%%%%%%%%%%%%%%%%%%%%%%%%%%%%%%%%%%%%%
In the recent past the phenomenon of ``inverse melting'', already hypothesized
by Tammann a century ago,\cite{Tammann} has been found experimentally in very
different materials, ranging from polymeric and colloidal compounds to 
high-$T_c$ superconductors, proteins, ultra-thin films, liquid crystals and
metallic alloys.\cite{RHKMM99,SDTJPC01,CACPS97,PVPNat03}
This kind of transition, includes, e.g., the solidification
of a liquid or the transformation of an amorphous solid into a crystal upon
heating. The reason for this counter intuitive process is that a phase usually
at higher entropic content happens to exist in very peculiar patterns such
that its entropy is decreased below the entropy of the phase usually
considered the most ordered one.  An example taking place in the widely
studied polymer P4M1P\cite{RHKMM99} is the one of a crystal state of
{\em higher} entropy that can be transformed into a fluid phase of {\em lower}
entropy on cooling, thus allowing, e.g., the melting of a crystal as the
temperature (or pressure) is decreased.  Inverse transitions, in their most
generic meaning (i.e. both thermodynamic or occurring by means of kinetic
arrest), have been observed between fluid and crystal phases,\cite{SDTJPC01}
between glass and crystal \cite{RHKMM99} and between fluid
and glass (``inverse freezing'').\cite{CACPS97}

A reentrance in the transition line
can be due both to the existence of a liquid phase with an entropy
lower than the one of the solid and/or because the liquid is more
dense than the solid (like in the water-ice transition).  When an
entropic ``inversion'' accounts for the phase transition, the
equilibrium transition line changes slope in a point where the entropy
of the fluid phase, $s_2$, becomes equal to the one of the solid, 
$s_1$, according to
the Clausius Clapeyron equation for first order phase transitions.
From this point a whole iso-entropic line, $\Delta s = s_2-s_1=0$, can
be continued both inside the solid and the liquid phases, as the
thermodynamic parameters (temperature and pressure for instance) are
varied.
This is a particularly interesting observation since, in the context
of glass formers, Kauzmann\cite{Kauz48} hypothesized a
transition to an ``ideal'' glass at the temperature at which $\Delta
s=0$, in order to avoid the paradox that an under-cooled liquid might
possess less entropy than the associated crystal at the same values of
the thermodynamic parameters.  From an experimental point of view the
Kauzmann temperature would be the temperature of the glass transition
(that is not a true phase transition because strictly kinetic in
origin) in an idealized adiabatic cooling procedure. Since the
astronomically long relaxation time needed to actually perform such an
experiment makes such a procedure unfeasible, the evidence in favor of
the existence of a thermodynamic glass transition mainly comes from
analytical and numerical investigations (see
e.g. Refs. \onlinecite{MPJCP,CPV1}).  The fact that a
$\Delta s=0$ line turns out naturally in the description of the
behavior of materials with inverse transition avoids, at least for
these substances, the Kauzmann paradox and breaks the connection
between $\Delta s=0$ extrapolation and the existence of an ideal
amorphous phase.\cite{SDTJPC01,RHKMM99}  

The aim of this paper is to study a simple mean-field model for the
inverse transition in a spin-glass, in order to heuristically
represent the inverse fluid-amorphous transition.
% yielding a stable
%thermodynamic solution everywhere in the parameter space and
%displaying a reentrance in the phase diagram.  
The model we consider displays quenched randomness as basic
ingredient.  We stress, however, that such a disorder is not necessary
to induce the spin-glass transition.  Truly relevant is the
frustration caused by it, i.e.  the unresolvable competition among
many similar states in the system evolution, but the source of
frustration can be of different nature, e.g. geometric. The {\em kind}
of frustration can actually lead to different amorphous phases (a
glass or a spin-glass) but it is not the quenched randomness the
discriminant factor.  Indeed, spin-glass models can be found without
quenched disorder\cite{MPR} as well as structural glass models with
quenched disorder (e.g. the ``discontinuous spin-glasses''
\cite{MPJCP} sharing the physical properties of a true glass rather
than of an amorphous magnet).

  We have been analyzing the
Blume-Emery-Griffiths-Capel (BEGC) model with quenched disorder using
the Full Replica Symmetry Breaking (FRSB) scheme of computation that
yields the exact stable thermodynamics.  The interested reader can
find details about the computation of the
thermodynamics of this kind of model in Refs. \onlinecite{CLPRL02,CLPRB04}.
It includes the Blume-Capel \cite{CapelPhys66} and the
Blume-Emery-Griffiths\cite{BEGPRA71} models, when the couplings are
ferromagnetic, and the Ghatak-Sherrington (GS) model,\cite{GSJPC77}
when the couplings are random variables and no biquadratic interaction
occurs (see Ref. \onlinecite{CLPRB04} for a more complete literature
report).  The model we have been extensively analyzing in the past is
the one with Ising spins ($S=\pm 1$) on a lattice gas (with site
occupation numbers $n=0,1$).  In that case the value associated with a
single site can be $1$, $0$ or $-1$, but zero has a double degeneracy
with respect to $1$ (or $-1$).  The original BEGC model consists,
instead, of spin-1 variables $S=1,0,-1$, with $1$ (or $-1$) being as
degenerate as $0$.  In their recent work\cite{SSPRL04,SS05} Schupper
and Shnerb introduced a generalization of the GS model to
theoretically represent the phenomenon of inverse freezing.  They
computed the phase diagram of such a model in the Replica Symmetric
(RS) approximation presenting evidence that a reentrance occurs if the
degeneracy of the magnetically interacting sites is larger than the
one of the holes.  Stimulated by their work we have been looking at
the phase diagram of the random BEGC model in terms of the variables
introduced by them, this time considering the thermodynamically stable
spin-glass obtained by means of the FRSB Ansatz, instead of the RS
approximation.  The aim of this work is (i) to check the validity of
the idea introduced in Ref.  \onlinecite{SSPRL04} in a non
pathological case and (ii) to determine the simplest model in which
the inverse freezing phenomenon is qualitatively well reproduced,
including the presence of latent heat.

The model  Hamiltonian we consider is 
\BEQ 
{\cal H}=\sum_{ij}J_{ij}S_iS_j+D\sum_{i=1}^NS_i^2
-\frac{K}{N}\sum_{i<j}S_i^2S_j^2
\label{Ham}
\EEQ 
where $S=1,0,-1$, $D$ is the crystal field, $J_{ij}$ are quenched
random variables (Gaussian) of mean zero and variance $1/N$.  The
parameter $K$ represents the strength of the biquadratic interaction.
%(corresponding to a particle-particle interaction in the lattice gas
%language). 
% Following the notation of Ref. \onlinecite{SSPRL04} 
We
call ${k}$ the degeneracy of the filled in sites of one type ($S=1$ or
$S=-1$) and ${l}$ the degeneracy of the empty sites ($S=0$). The
relevant parameter is $r={k}/{l}$.\cite{SSPRL04} 
When $r=1$ the {spin-1} model is
obtained. If, furthermore, $K=0$, the model is the GS one.  When,
otherwise, $r=1/2$ and $D\to\mu= -D$ the lattice gas formulation of
Ref.  \onlinecite{ANSJPF96} is recovered, for which no reentrance was
observed.\cite{CLPRL02,CLPRB04} 
%A reentrance was there, instead, as
%the expansion for small values of the order parameter identifying the
%frozen phase was performed around the tricritical
%point.\cite{MSJPC85,CLPRB04} We stress, though, that this reentrance
%is not an artifact of the naive RS Ansatz but only a consequence of
%the finiteness of the order of the expansion (whatever Ansatz is
%employed) as one goes away from the tricritical point.

Schupper and Shnerb bring about the idea that a larger degeneracy of
the interacting sites yields a qualitative change of the phase
diagram up to develop a reentrance in the $T-D$ phase diagram, thus
allowing for a phase transition from the paramagnetic (PM) to the
spin-glass (SG) phase as temperature is {\em increased}.  Their statement is
that, if the ratio $r$ is large enough, the phase diagram changes so
much up to display a reentrance in the $T-D$ plane, as they show in
the RS case.\cite{SSPRL04} This solution turns out to be
unphysical in any phase that is not the PM one\cite{Almeida}
and the shape of the transition line and of the SG spinodal line
might then be sensitive to the thermodynamic instability intrinsic in
such approximation.  Therefore, moving to the right RSB scheme of
computation, there is no guarantee that the first order transition and
the spin glass spinodal lines would remain the same. On the other
hand, not even the certainty exists that the first order phase
transition line between the PM and the SG phase
computed in the RS scheme has to be displaced with respect to the
PM/SG(FRSB) transition line up to developing a reentrance. 

\begin{figure}[t!]
\includegraphics[width=.49\textwidth]{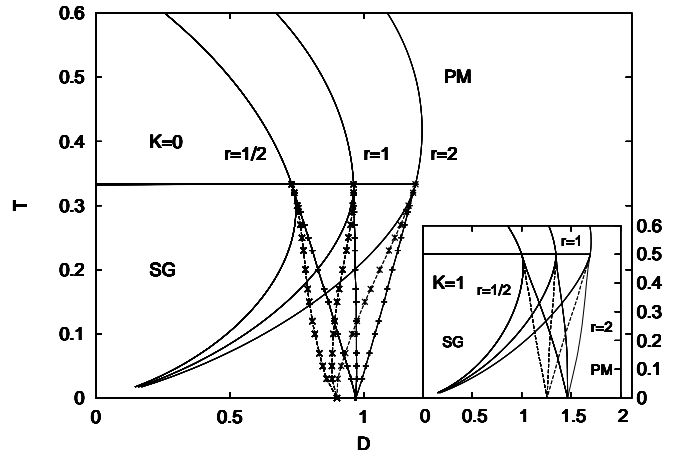}
\caption{The $D$-$T$ phase diagram in absence of biquadratic
  interaction.  Three models with different behaviors are plotted:
  $r=1/2$, $1$, $2$.  For each model three curves are represented,
  each departing from the same tricritical point: the full curve on
  the left is the spinodal of the PM phase, the dashed one in the
  middle is the first order transition line and the right one is the
  SG spinodal line.  The group of three curves on the left are for the
  Ising spins on lattice gas ($r=1/2$, $T_c=1/3$, $D_c=0.73105$). The
  group of curves in the middle represent the lines of the GS model
  ($r=1$, $D_c=0.96210$). The curves on the right correspond to the
  $r=2$ model ($D_c=1.19315$).  In the inset the same diagram is plotted
  when $K=1$ (attractive biquadratic interaction).}
\label{fig:phdiK0}
\end{figure}

\begin{figure}[t!]
\includegraphics[width=.49\textwidth]{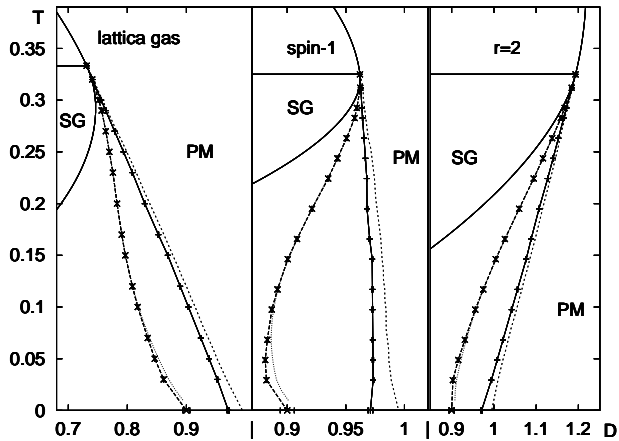}
\caption{$D$-$T$ phase diagrams at $K=0$
  of the Ising spin glass on lattice gas (left), of the GS model (center) and
  of the random BEGC model with $r=2$ (right).  Both the RS and the FRSB
  solution (the latter with error bars) are plotted. In the first case no
  reentrance takes place, disregarding the approximation. A reentrance occurs,
  instead, below the tricritical point in the spin-1 model along the first
  order transition line ans a second reentrance seems to be there for
  $T<0.03$. In the latter model 
  the reentrance occurs above the tricritical point.}
\label{fig:rdet}
\end{figure}

We discuss the physically stable solution for both the GS
model\cite{GSJPC77,MSJPC85} ($K=0$) and the model with attractive biquadratic
interaction\cite{CLPRB04} ($K/J=1$), whose phase diagrams are plotted in Fig.
\ref{fig:phdiK0}, and we compare it with the RS results (see Fig.
\ref{fig:rdet}).  We study the behavior of the phase diagram for (a) the
lattice gas case ($r=1/2$), for which no inverse transition occurs anywhere in
the parameter space; (b) the spin-1 case ($r=1$), where the first order
transition line displays a reentrance soon below the tricritical temperature;
(c) the generalized cases as $r>1$, in particular we plot the results of the
model with variables taking values $S=\{1,1,0,-1,-1\}$ for which the reentrance
takes place above the tricritical point, along the second order phase
transition line.

Having a model with variables displaying a relative degeneracy $r$
($D=D_r$), in order to describe the partition function of another model
whose variables have degeneracy $r'$ 
it is enough to vary the crystal field as
$ 
D_{r' }+ T\log r' = D_{r} +
T\log r
$. 
This does not hold, however, for the state
functions obtained deriving the thermodynamic potential 
with respect to the temperature
(e.g. entropy and internal energy) 
that will, instead, receive contributions from additional terms.
  Identifying $D_{1/2}\equiv -\mu$ one can recover the case of
magnetic spins on a lattice gas of chemical potential
$\mu$.\cite{ANSJPF96,CLPRL02}

\begin{figure}[t!]
\includegraphics[width=.49\textwidth]{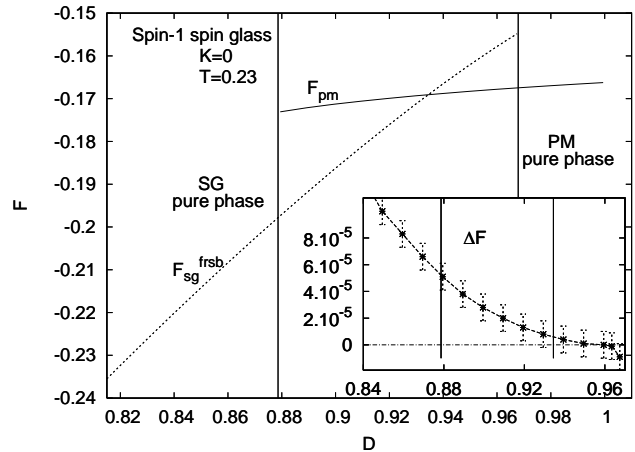}
\caption{The free energy $F_{\rm pm}$ and $F_{\rm sg}^{\rm frsb}$
  versus the crystal field $D$ at $T=0.23$ in the GS model
  ($K=0$). The left side vertical line is at the spinodal point of the
  PM phase, the right one at the spinodal point of the SG phase. 
  The first order transition occurs at $D_{1^{st}}=0.9344$. In
  the inset $\Delta F=F_{\rm sg}^{\rm frsb}-F_{\rm
  sg}^{\rm rs}$ is displayed. At this temperature the two functions
  merge very near to the tricritical point (right side vertical line).}
\label{fig:F_D}
\end{figure}

The analysis leads us to the conclusion that the transition lines are
not very much dependent on the Ansatz used to compute the quenched
average of the free energy. Actually, for not extremely low $T$, the
first order transition lines yielded by the RS and the FRSB Ansatz
coincide down to the precision of our numerical evaluation of the FRSB
antiparabolic Parisi equation,\cite{ParisiJPA80}
 whereas they slightly differ at very
low temperature (see Fig.  \ref{fig:rdet}). For what concerns the
spinodal lines, the RS ones are shifted by a small amount inside the
pure PM phase.  In order to clarify this point the behavior of the
free energy vs. D at fixed temperature ($T=0.23$) is displayed in
Fig. \ref{fig:F_D} for the spin-1 model in absence of biquadratic
interaction.  In the inset the difference between the FRSB and the RS
free energies ($\Delta F\equiv F_{\rm frsb}-F_{\rm rs}$) is
plotted. In the coexistence region we have a subregion where the
stable phase is PM (r.h.s. of fig. \ref{fig:F_D}),
i.e. $F_{\rm pm} < F_{\rm sg}$, and a complementary subregion where
$F_{\rm sg}<F_{\rm pm}$. The free energy of the stable spin glass
phase is, of course, the FRSB one. In the SG phase any
approximation of this free energy yielded by means of a finite number
of RSB leads to a {\em{lower} } value (and an unstable
phase),\cite{ParisiJPA80} and we can see in the inset that 
%($D_{1^{st}}=0.9344$)
 $F_{\rm frsb}>F_{\rm rs}$.
 The two free energies merge
 around the first order transition (compatibly with the
numerical uncertainty).
%  The free energy values computed at
%$T=0.23$ have a numerical error of $10^{-5}$ and
At $T=0.23$ in the phase
coexistence region the two Ansatz yield very similar values (as
opposed, e.g., to the behavior deep in the SG pure phase), so that it
is not possible to determine exactly the merging point of $F_{\rm
rs}(D)$ and $F_{\rm frsb}(D)$. 
%also looking at the results at lower temperature
(see Fig. \ref{fig:rdet}), 
One can infer, however, that it remains above $D_{1^{st}}(T)$.

We show the $D$-$T$ phase diagrams for $K=0$ and $K=1$ in Fig.
\ref{fig:phdiK0} and inset, at different values of $r$.  The diagram
of the model with Ising spins on lattice gas is here represented as a
function of the parameter $D=-\mu$, in order to simplify the
comparison with the spin-1 model and the $r>1$ cases.  We find
that the reentrance in the $D$-$T$ plane is present already in the
spin-1 GS model.  As a consequence this implies that there is no need
for the intuition of Ref.  \onlinecite{SSPRL04} in order to
have a model for inverse freezing from low temperature liquid to high
temperature amorphous solid.  This is at difference with the
liquid-crystal inverse transition (``inverse melting'') for the
description of which the original Blume-Capel model is not adequate
and $r>1$ is needed.\cite{SSPRL04,SS05}

The slope of a first order line is given by the Clausius-Clapeyron
equation.  For the BEGC model it can be written in terms of the crystal
field $D$ (playing the role of a chemical potential), instead of the
pressure that is not defined in our model:
\BEQ
\frac{d D}{d T}=\frac{s_{\rm pm}-s_{\rm sg}}{\rho_{\rm pm}-\rho_{\rm sg}}
=\frac{\Delta s}{\Delta\rho}
\label{CC}
\EEQ This formula is valid for any $r$. We stress that in passing from
$r$ to $r'$ also the entropy changes of a term $\rho \log r/r'$, in
agreement with the crystal field shift given above.
Going down along the transition line, as $\Delta s$ changes sign the
slope becomes positive. The $\Delta s=0$ point is called a Kauzmann
locus.\cite{SDTJPC01}

\begin{figure}[t!]
\includegraphics[width=.49\textwidth]{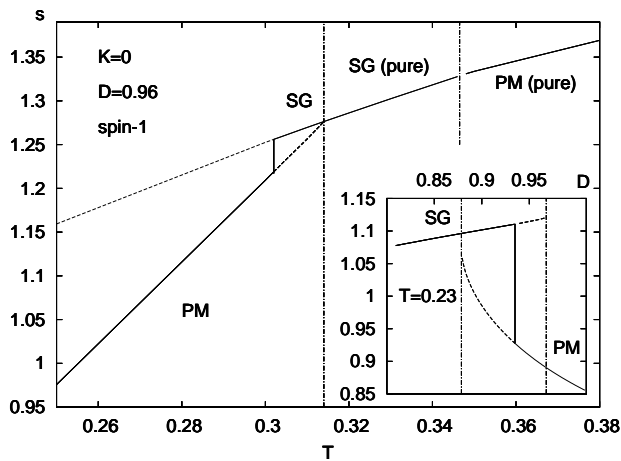}
\caption{Entropy vs. $T$ at the crystal field value of $D=0.96$ for
the GS model. $T_{1^{st}}=0.302$. The two vertical dashed lines are
the SG ($T_{\rm SG}=0.314$) and PM ($T_{\rm PM}=0.3465$) spinodal
lines.  In the inset $s(D)$ is plotted at $T=0.23$.}
%($D_{\rm SG}=0.8786$, $D_{1^{st}}=0.9349$, $D_{\rm PM}=0.9679$). }
\label{fig:entro}
\end{figure}

Looking at the phase diagram in the spin-1 case one can observe that
the RS first order phase transition line also displays a second
turning as the temperature becomes lower and lower (see
Fig. \ref{fig:rdet}).  Such a turning is less evident as the FRSB
solution is considered but it does not disappear. 
%(even though the
%evaluation of the critical point at $T=0$ is less accurate than for
%$T>0$, the bending is more pronounced than the numerical error). 
 The
low temperature turning is not there, instead,
 for $r=2$. We notice also that in
this last case the reentrance is already in the second order phase
transition line and both a first and a second order inverse freezing
transitions are possible.  In particular a transition with exchange of
latent heat can occur exclusively in the {\rm inverse} order.  In
Fig. \ref{fig:entro} the behavior of the entropy as a function of the
temperature is shown across an inverse transition (as a function of
the crystal field $D$ in the inset) for the spin-1 model.  The entropy
of the PM phase {\em below} the first order transition line is {\em
smaller} than the entropy of the SG: heating the system the paramagnet
becomes an amorphous magnet (i.e. ``freezes'') acquiring latent heat
from the heat bath.

Introducing a biquadratic interaction term and varying it from
attractive to repulsive the situation does not change much (see
e.g. inset of Fig. \ref{fig:phdiK0}). 
For any value of  $K$ no
reentrance of the phase transition line occurs in the $D$-$T$ phase
diagram of the lattice gas model whereas it is {\em always} there for
the spin-1 model. The only consequence of reducing $K$
 is that the area of the phase coexistence region is
reduced (the tricritical temperature tends to zero as $K\to-\infty$).

In conclusion we have shown that the Ghatak-Sherrington model,
 i.e. the Blume-Capel model with quenched disordered
magnetic interactions, computed in the exact FRSB Ansatz,
undergoes the inverse freezing phenomenon acquiring
latent heat from the heat bath as the paramagnet becomes a spin-glass.
Many other
models can be built starting from this one, introducing an attractive
or repulsive biquadratic interaction (the last term in Hamiltonian
(\ref{Ham})) and/or tuning the relative degeneracy of the value $S=0$
and $S^2=1$ of the spin variable (the Schupper-Shnerb
``spin''\cite{SSPRL04}) but the GS one already contains 
all features needed to qualitatively
represent the experimental results.

\acknowledgments LL thanks S. Rastogi for providing very 
interesting bibliography and L. Palatella and L. Angelani
for stimulating discussions.

\end{document}